\documentclass[twocolumn,twoside,slac_two]{revtex4}
\usepackage{graphicx}
\usepackage{fancyhdr}
\begin{document}

\title{Ancient Pulsar Wind Nebulae in light of recent GeV and TeV observations}

%

\author{O. Tibolla, K. Mannheim, D. Els\"asser}
\affiliation{ITPA, Universit\"at W\"urzburg, Campus Hubland Nord, Emil-Fischer-Str. 31 D-97074 W\"urzburg, Germany}
\author{S. Kaufmann}
\affiliation{Landessternwarte, Universit\"at Heidelberg, K\"onigstuhl, D 69117 Heidelberg, Germany}
\author{IN MEMORY OF OKKIE DE JAGER}

\begin{abstract}

In a Pulsar Wind Nebula (PWN), the lifetime of inverse Compton emitting electrons exceeds the lifetime of its progenitor pulsar, but it exceeds also the age of the electrons
that emit via synchrotron radiation; i.e. during the evolution of the PWN, it can remain bright in IC, whereas its GeV-TeV gamma-ray (for 10$^5$-10$^6$ years) flux remains high for timescales much larger than the Pulsar lifetime and the PWN visible in X-rays.  The shell-type remnant of the supernova explosion in which the pulsar was formed also has a much shorter lifetime. In this scenario, the magnetic field in the cavity induced by the wind of the progenitor star plays a crucial role. This is in line with the discovery of several unidentified sources in the TeV gamma-ray band without X-ray counterparts. Moreover, the consequences are important also in order to reinterprete the detection of starburst galaxies in the TeV gamma-ray band considering a leptonic origin of the gamma-ray signal.

\end{abstract}

\maketitle

\thispagestyle{fancy}


\section{Pulsar Wind Nebulae}

Pulsar wind nebulae (PWN) represent a unique laboratory for the highly efficient acceleration of particles to ultra-relativistic energies.
Early ideas about particle acceleration by electromagnetic waves emitted by pulsars due to \cite{1} led to the discovery by \cite{2} that a significant fraction of the spin-down power of a pulsar such as Crab is dissipated through a relativistic wind with in situ acceleration of particles at its termination shock, since the cooling time scales of the synchrotron-emitting electrons are very short in X-rays. Further developments led to the model of \cite{3} involving a low level of magnetization at the shock, raising the question of the transition from high to low sigma since the launch of the wind consisting of pairs and Poynting flux involves a high magnetization 

\begin{equation}
\sigma = \frac{B^2}{4\pi \rho c^2\gamma^2} \gg 1
\end{equation}

where $\rho$ denotes the density and $\gamma$ the Lorentz factor of the wind electrons and positrons. Recently, MHD simulations of such winds were performed by \cite{4} \cite{dz}. Central of our current understanding of PWN is that they are equatorial and highly relativistic, showing features such as backflows and jets. This might provide clues to their observational identification, since energy release presumably due to reconnection can lead to flares from the wind, considering relativistic transport effects and beaming, as recently observed in Crab. Moreover, the MHD models provide clues for predicting the temporal evolution of PWN, and to bring this into accordance with the observations of a large sample of putative PWN of different ages.

\section{Ancient Pulsar Wind Nebulae}

\begin{figure}
  \vspace{5mm}
  \centering
  \includegraphics[width=\columnwidth]{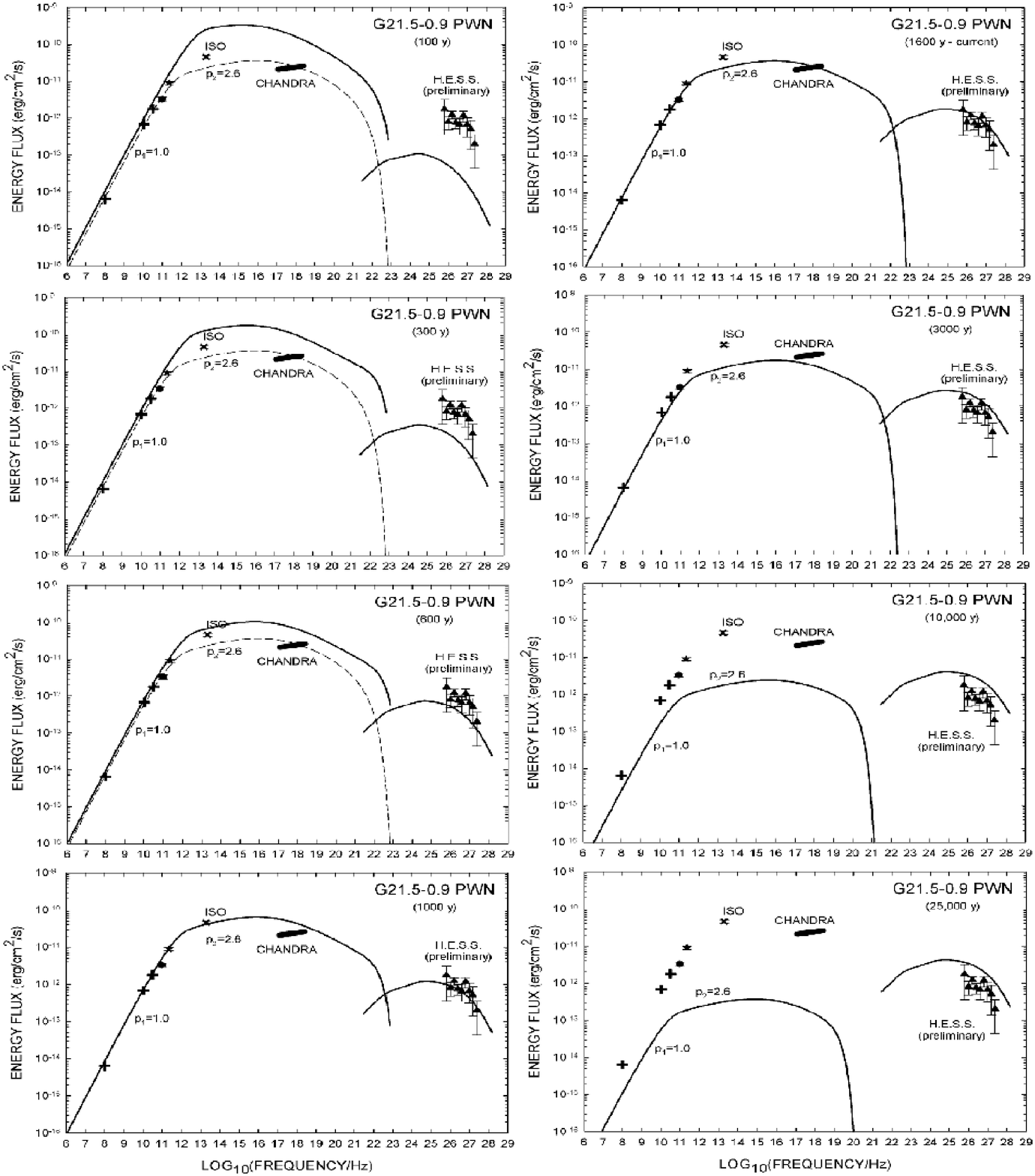}
  \caption{Representation of the Spectral Energy Distribution of G21.5-0.9 during the first 25 kyrs from \cite{10}; this example describes efficiently the temporal behavior of synchrotron and IC components}
  \label{fig1}
 \end{figure}

Although they are often detected as non-thermal X-ray sources, an evolved PWN can indeed lead to a fairly bright $\gamma$-ray source without any lower energy counterpart \cite{5} \cite{6}. The key issue is that the low energy synchrotron emission, where $\tau_E$ is the synchrotron emitting lifetime of TeV $\gamma$-ray emitting leptons with energy E, depends on the internal PWN magnetic field \cite{3} which may vary as a function of time, following $\tau_E \propto t^{2 \alpha}$ if $B(t) \propto t^{-\alpha}$, where $\alpha$ is the power-law index of the decay of the average nebular field strength, whereas the VHE emission depends on the CMB radiation field, which is constant on timescales relevant for PWN evolution. MHD simulations of composite SNRs \cite{7} find that $\alpha=1.3$ until the passage of the reverse shock; after it is expected that field decay would continue for much longer time since expansion continues even after the passage of the reverse shock.
From hydrodynamic simulations, an expression for the time of the reverse shock passage (i.e. the return time of the reverse shock to the origin) has been given \cite{8} as

\begin{equation}
T_R = 10~\mathrm{kyr} \left( \frac{\rho_{ISM}}{10^{-24}\mathrm{\frac{g}{cm^{3}}}} \right)^{-\frac{1}{3}}   \left(\frac{M_{ej}}{10 M_{\odot}}\right)^{\frac{3}{4}}   \left(\frac{E_{ej}}{10^{51}\mathrm{erg} }\right)^{-\frac{2}{3}}
\end{equation}

where the first term is the density of the ISM, the second the ejecta mass during the SNR explosion and the third the SNR blast wave energy. The stellar wind of a high-mass star can blow a cavity around the progenitor star \cite{9} with relatively low ISM density, so that $T_R >> 10~\mathrm{kyr}$. In such a case it is expected that $B(t)$ can decay as $t^{- \alpha}$, until the field is low enough for the X-ray flux to drop below the typical sensitivity levels. As a result, in a scenario where the magnetic field decays as a function of time, the synchrotron emission will also fade as the PWN evolves. The reduced synchrotron losses for high-energy electrons for such a scenario will then lead to increased lifetimes for these leptonic particles. For timescales shorter than the inverse-Compton lifetime of the electrons ($t_{IC} \propto 1.2 \times 10^6 (E_e/ 1 \mathrm{TeV})^{-1}$ years), this will result in an accumulation of VHE electrons which will also lead to an increased gamma-ray production due to up-scattering of CMB photons. Such accumulation of very-high energy electrons in a PWN has indeed been seen in many TeV PWNe (such as HESS J1825-137 \cite{1825}). To summarize, during their evolution PWN may appear as gamma-ray sources with only very faint low-energy counterparts and this may represent a viable model for many unidentified TeV sources. This effect can be clearly seen in Fig. \ref{fig1}, shown by Okkie de Jager in SciNeGHE 2008 conference \cite{10}.

\section{Link between PWN timescales and Galactic latitudes}

There are numerous observational hints pointing to the fact that the interstellar magnetic field in the Milky Way has field strength of the order of several microgauss in the inner Galactic disc, and presumably drops to tenths of microgauss at higher latitudes \cite{11} \cite{12}.
Although neglected in previous studies, this magnetic field of the ISM into which the Pulsar Wind Nebula is expanding must have some effect onto the morphology and time evolution of the PWN.
As the magnetic field in the PWN drops rather rapidly over time, as $t^{-1.3}$ , it becomes comparable to the surrounding ISM field at timescales roughly comparable to the transition timescale into the post-Sedov phase. However, as noted above, different PWN will be located in environments where the ambient field strength is different by factors of order unity, depending rather strictly on the location of the PWN in the Galaxy. The respective timescales for the field to become comparable to the ISM field will therefore differ by a factor of $\sim 3$ for a PWN at the inner disc compared to one in the Galactic halo. Detailed studies of the effects on PWN morphology and evolution are highly demanding and yet challenging, owing to the limited resolution of high energy instruments and the relative scarcity of identified PWN.

\section{Two other important corollaries}

\subsection{TeV unidentified sources}

Ancient PWNe are a natural explanation for TeV unidentified sources \cite{7}. In fact $\sim 50$\% of TeV Galactic sources are still formally unidentified \cite{13}; almost all (among them, only HESS J0632+057 \cite{mon}, that is considered a $\gamma$-ray binary candidate, and HESS J1943+213 \cite{hbl}, considered a HBL candidate but recently suggested to be more likely a PWN \cite{1943pwn}, are point-like) are extended objects with angular sizes ranging from approximately 3 to 18 arc minutes (but of few more extended exceptions, such as HESS J1841-055 \cite{unids}), lying close to the Galactic plane (suggesting a location within the Galaxy). In each case, the spectrum of the sources in the TeV energy range can be characterized as a power-law with a differential spectral index in the range 2.1 to 2.5. The general characteristics of these sources (spectra, size, and position) are similar to previously identified galactic VHE sources (e.g. PWNe), however these sources have so far no clear counterpart in lower-energy wavebands. This scenario is particularly suitable for several sources that, even after deep MWL observations, lack any plausible X-ray counterparts (such as HESS J1507-622 \cite{6}, HESS J1427-608, HESS J1708-410 \cite{unids} and HESS J1616-508 \cite{surv} \cite{suz}). On the other hand, several unidentified sources have already been identified as PWNe after their discovery (such as HESS J1857+026 \cite{unids} or HESS J1303-631 \cite{1303}) and in sources that have several plausible counterparts, the PWNe contribution can hardly be avoided (such as hot spot B in HESS J1745-303 \cite{1745} or HESS J1841-055).

\begin{figure}
  \vspace{5mm}
  \centering
  \includegraphics[width=\columnwidth]{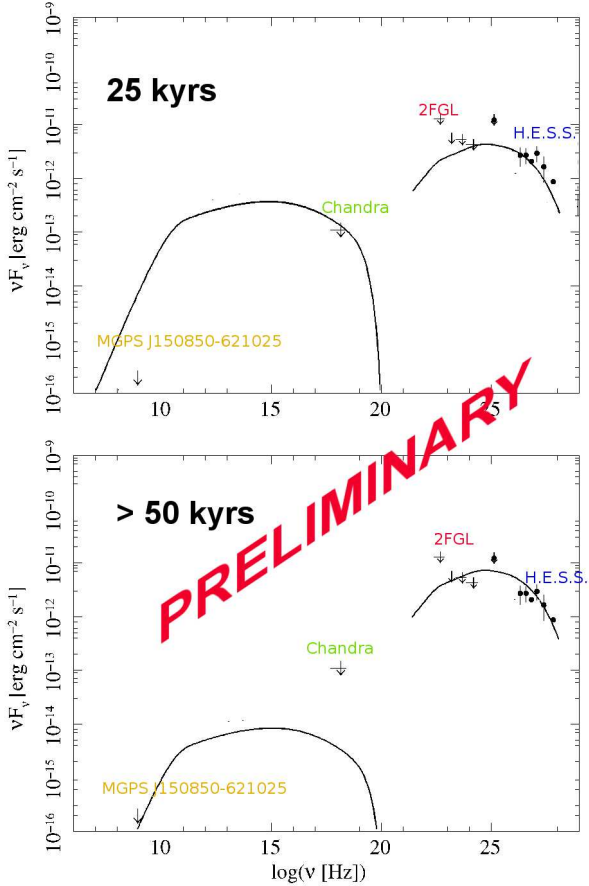}
  \caption{Preliminary representation of the Spectral Energy Distribution of HESS J1507-622 assuming initial conditions similar to the G21.5-0.9/PSR J1833-1034 system; the radio information is taken from \cite{MGPS}, the Fermi-LAT points from \cite{2FGL}, the Chandra and H.E.S.S. points from \cite{6}.}
  \label{fig3}
 \end{figure}

Fig. \ref{fig3} shows the example of HESS J1507-622: this VHE source does not have any SNR as a possible counterpart and does not show any coincident or close-by pulsar \cite{6}; as said, it is expected that the pulsar already spun-down and that the X-ray nebula faded away below the sensitivity of current X-ray instruments, however it requires an arbitrary choice of the initial conditions in order to model this source: in the shown example we assume to have conditions similar to the G21.5-0.9/PSR J1833-1034 system.

\subsection{Starburst galaxies}

 \begin{figure}
  \vspace{5mm}
  \centering
  \includegraphics[width=\columnwidth]{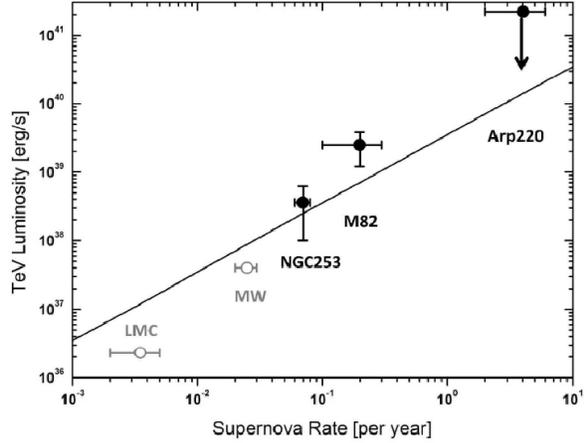}
  \caption{This figure taken from \cite{15} shows the predicted TeV gamma-ray luminosity from PWN (solid line) for different core-collapse supernova rates.  Note that the points shown for LMC and the Milky Way are extrapolations of measurements obtained at Fermi energies.}
  \label{fig2}
 \end{figure}

Another important consequence of these long-living gamma-ray sources regards starburst galaxies: in fact it has been recently shown \cite{15} that PWNe are important and not negligible in explaining the TeV emission detected from NGC 253 \cite{16} and from M82 \cite{17}. In fact, even if M82 and NGC253 show a much harder GeV gamma-ray spectrum than the Milky Way (in line with the high gas density in the star forming regions (SFRs) and a cosmic-ray origin of the gamma rays), diffusive-convective escape of the cosmic rays from the SFRs should lead to a steepening of the cosmic-ray induced gamma-ray spectrum above $\simeq$ 10 GeV, in which case the cosmic rays would fall short in explaining the TeV luminosities; on the other hand PWNe associated with core-collapse supernovae in SFRs can readily explain the observed high TeV luminosities. The proof of this could arrive from deeper gamma-ray observations on other galaxies, as shown in Fig. \ref{fig2}. 

Moreover also neutrino observations could reveal if the PWN play a role in accelerating cosmic ray protons and ions; the highly magnetized pulsar equatorial wind may also give rise to non-negligible acceleration of cosmic rays by this source class.  Protons and ions, swept into the plerion by the reverse shock, could be accelerated efficiently at the ultrarelativistic pulsar wind shocks \cite{neu1} \cite{neu2}. This scenario could possibly be tested by  searching for high-energy neutrinos from galactic PWNs like the Crab.
Until now, such measurements have turned up only null-results. However, a stacking analysis of candidate sources could significantly enhance the neutrino telescope sensitivity, thus resulting
in a much more decisive test of this scenario.
Moreover recently strong upper limits have been given by means of Ice Cube (e.g. \cite{18}), limits close to the theoretical predictions.

\bigskip 
\begin{acknowledgments}

The idea of presenting this on-going job came as a small sign to honor the memory of a great scientist and a wonderful person: Okkie de Jager.
Okkie, working with you was a great honor we will miss you very much! Thank you also for encouraging to put forward this research. \\

BMBF contract 05A08WW1 is also acknowledged.

\end{acknowledgments}

\bigskip 

\end{document}